\documentstyle[twoside,psfig]{article}

\catcode`\@=11
\long\def\@makefntext#1{
\protect\noindent \hbox to 3.2pt {\hskip-.9pt  
$^{{\eightrm\@thefnmark}}$\hfil}#1\hfill}		

\def\thefootnote{\fnsymbol{footnote}}
\def\@makefnmark{\hbox to 0pt{$^{\@thefnmark}$\hss}}	
	
\def\ps@myheadings{\let\@mkboth\@gobbletwo
\def\@oddhead{\hbox{}
\rightmark\hfil\eightrm\thepage}   
\def\@oddfoot{}\def\@evenhead{\eightrm\thepage\hfil
\leftmark\hbox{}}\def\@evenfoot{}
\def\sectionmark##1{}\def\subsectionmark##1{}}

\oddsidemargin=\evensidemargin
\renewcommand{\thefootnote}{\fnsymbol{footnote}}

\newcounter{sectionc}\newcounter{subsectionc}\newcounter{subsubsectionc}
\renewcommand{\section}[1] {\vspace{12pt}\addtocounter{sectionc}{1} 
\setcounter{subsectionc}0 \setcounter{subsubsectionc}0 \noindent 
	{\tenbf\thesectionc. #1}\par\vspace{5pt}}
\renewcommand{\subsection}[1] {\vspace{12pt}\addtocounter{subsectionc}{1} 
	\setcounter{subsubsectionc}0 \noindent 
	{\bf\thesectionc.\thesubsectionc. {\kern1pt \bfit #1}}\par\vspace{5pt}}
\renewcommand{\subsubsection}[1] {\vspace{12pt}\addtocounter{subsubsectionc}{1}
	\noindent{\tenrm\thesectionc.\thesubsectionc.\thesubsubsectionc.
	{\kern1pt \tenit #1}}\par\vspace{5pt}}
\newcommand{\nonumsection}[1] {\vspace{12pt}\noindent{\tenbf #1}
	\par\vspace{5pt}}

\newcounter{appendixc}
\newcounter{subappendixc}[appendixc]
\newcounter{subsubappendixc}[subappendixc]
\renewcommand{\thesubappendixc}{\Alph{appendixc}.\arabic{subappendixc}}
\renewcommand{\thesubsubappendixc}
	{\Alph{appendixc}.\arabic{subappendixc}.\arabic{subsubappendixc}}

\renewcommand{\appendix}[1] {\vspace{12pt}
        \refstepcounter{appendixc}
        \setcounter{figure}0 
        \setcounter{table}0 
        \setcounter{lemma}0 
        \setcounter{theorem}0 
        \setcounter{corollary}0 
        \setcounter{definition}0 
        \setcounter{equation}0 
        \renewcommand{\thefigure}{\Alph{appendixc}.\arabic{figure}}
        \renewcommand{\thetable}{\Alph{appendixc}.\arabic{table}}
        \renewcommand{\theappendixc}{\Alph{appendixc}}
        \renewcommand{\thelemma}{\Alph{appendixc}.\arabic{lemma}}
        \renewcommand{\thetheorem}{\Alph{appendixc}.\arabic{theorem}}
        \renewcommand{\thedefinition}{\Alph{appendixc}.\arabic{definition}}
        \renewcommand{\thecorollary}{\Alph{appendixc}.\arabic{corollary}}
        \renewcommand{\theequation}{\Alph{appendixc}.\arabic{equation}}
        \noindent{\tenbf Appendix \theappendixc #1}\par\vspace{5pt}}
\newcommand{\subappendix}[1] {\vspace{12pt}
        \refstepcounter{subappendixc}
        \noindent{\bf Appendix \thesubappendixc. {\kern1pt \bfit #1}}
	\par\vspace{5pt}}
\newcommand{\subsubappendix}[1] {\vspace{12pt}
        \refstepcounter{subsubappendixc}
        \noindent{\rm Appendix \thesubsubappendixc. {\kern1pt \tenit #1}}
	\par\vspace{5pt}}

\newcommand{\textlineskip}{\baselineskip=13pt}
\newcommand{\smalllineskip}{\baselineskip=10pt}

\def\eightcirc{
\begin{picture}(0,0)
\put(4.4,1.8){\circle{6.5}}
\end{picture}}
\def\eightcopyright{\eightcirc\kern2.7pt\hbox{\eightrm c}}

\def\abstracts#1#2#3{{
	\centering{\begin{minipage}{4.5in}\footnotesize\baselineskip=10pt
	\parindent=0pt #1\par 
	\parindent=15pt #2\par
	\parindent=15pt #3
	\end{minipage}}\par}} 

\newcommand{\bibit }{\nineit}
\newcommand{\bibbf}{\ninebf}
\renewenvironment{thebibliography}[1]
	{\frenchspacing
	 \ninerm\baselineskip=11pt
	 \begin{list}{\arabic{enumi}.}
	{\usecounter{enumi}\setlength{\parsep}{0pt}
	 \setlength{\leftmargin 12.7pt}{\rightmargin 0pt} 
	 \setlength{\itemsep}{0pt} \settowidth
	{\labelwidth}{#1.}\sloppy}}{\end{list}}

\newcounter{itemlistc}
\newcounter{romanlistc}
\newcounter{alphlistc}
\newcounter{arabiclistc}

\newcommand{\fcaption}[1]{
        \refstepcounter{figure}
        \setbox\@tempboxa = \hbox{\footnotesize Fig.~\thefigure. #1}
        \ifdim \wd\@tempboxa > 5in
           {\begin{center}
        \parbox{5in}{\footnotesize\smalllineskip Fig.~\thefigure. #1}
            \end{center}}
        \else
             {\begin{center}
             {\footnotesize Fig.~\thefigure. #1}
              \end{center}}
        \fi}

\newcommand{\tcaption}[1]{
        \refstepcounter{table}
        \setbox\@tempboxa = \hbox{\footnotesize Table~\thetable. #1}
        \ifdim \wd\@tempboxa > 5in
           {\begin{center}
        \parbox{5in}{\footnotesize\smalllineskip Table~\thetable. #1}
            \end{center}}
        \else
             {\begin{center}
             {\footnotesize Table~\thetable. #1}
              \end{center}}
        \fi}

\def\@citex[#1]#2{\if@filesw\immediate\write\@auxout
	{\string\citation{#2}}\fi
\def\@citea{}\@cite{\@for\@citeb:=#2\do
	{\@citea\def\@citea{,}\@ifundefined
	{b@\@citeb}{{\bf ?}\@warning
	{Citation `\@citeb' on page \thepage \space undefined}}
	{\csname b@\@citeb\endcsname}}}{#1}}

\newif\if@cghi
\def\cite{\@cghitrue\@ifnextchar [{\@tempswatrue
	\@citex}{\@tempswafalse\@citex[]}}
\def\citelow{\@cghifalse\@ifnextchar [{\@tempswatrue
	\@citex}{\@tempswafalse\@citex[]}}
\def\@cite#1#2{{$\null^{#1}$\if@tempswa\typeout
	{IJCGA warning: optional citation argument 
	ignored: `#2'} \fi}}

\def\pmb#1{\setbox0=\hbox{#1}
	\kern-.025em\copy0\kern-\wd0
	\kern.05em\copy0\kern-\wd0
	\kern-.025em\raise.0433em\box0}

\def\fnt#1#2{\footnotetext{\kern-.3em
	{$^{\mbox{\scriptsize #1}}$}{#2}}}

\def\thefootnote{\fnsymbol{footnote}}
\def\@makefnmark{\hbox to 0pt{$^{\@thefnmark}$\hss}}	
	
\def\ps@myheadings{%
    \let\@oddfoot\@empty\let\@evenfoot\@empty
    \def\@evenhead{\slshape\leftmark\hfil}
    \def\@oddhead{\hfil{\slshape\rightmark}}
    \let\@mkboth\@gobbletwo
    \let\sectionmark\@gobble
    \let\subsectionmark\@gobble
    }
\font\tenrm=cmr10
\font\tenit=cmti10 
\font\tenbf=cmbx10
\font\bfit=cmbxti10 at 10pt
\font\ninerm=cmr9
\font\nineit=cmti9
\font\ninebf=cmbx9
\font\eightrm=cmr8

\def\qed{\hbox{${\vcenter{\vbox{			
   \hrule height 0.4pt\hbox{\vrule width 0.4pt height 6pt
   \kern5pt\vrule width 0.4pt}\hrule height 0.4pt}}}$}}

\renewcommand{\thefootnote}{\fnsymbol{footnote}}  

\def\beq{\begin{eqnarray}}
\def\eeq{\end{eqnarray}}
\def\kB{k_{\rm B}}
\def\lP{\ell_{\rm P}}
\def\lC{\ell_{\rm C}}
\def\tP{t_{\rm P}}
\def\mP{m_{\rm P}}
\def\Hz{{\rm Hz}}
\def\K{{\rm K}}
\def\gw{{\rm gr}}

\def\m{{\rm m}}
\def\g{{\rm g}}

\begin{document}
\setlength{\textheight}{7.7truein}  

\thispagestyle{empty}

\markboth{\protect{\footnotesize\it Decoherence and gravitational
backgrounds}}{\protect{\footnotesize\it Decoherence and gravitational
backgrounds}}

\normalsize\textlineskip

\setcounter{page}{1}

\vspace*{0.88truein}

\centerline{\bf DECOHERENCE AND GRAVITATIONAL BACKGROUNDS}
\vspace*{0.37truein}
\centerline{\footnotesize Serge REYNAUD 
\footnote{mailto:reynaud@spectro.jussieu.fr; 
http://www.spectro.jussieu.fr/Vacuum} ,
Brahim LAMINE, Astrid LAMBRECHT}
\baselineskip=12pt
\centerline{\footnotesize\it Laboratoire Kastler Brossel
\footnote{Laboratoire du CNRS, de l'Ecole Normale Sup\'{e}rieure 
et de l'Universit\'{e} Pierre et Marie Curie.} , UPMC case 74 }
\baselineskip=10pt
\centerline{\footnotesize\it Campus Jussieu, F-75252 Paris Cedex 05, France}
\vspace*{10pt}
\centerline{\footnotesize Paulo MAIA NETO}
\baselineskip=12pt
\centerline{\footnotesize\it Instituto de F\'{\i}sica, UFRJ, Caixa Postal 68528}
\baselineskip=10pt
\centerline{\footnotesize\it 21945-970 Rio de Janeiro, Brazil}
\vspace*{10pt}
\centerline{\footnotesize Marc-Thierry JAEKEL}
\baselineskip=12pt
\centerline{\footnotesize\it Laboratoire de Physique Th\'{e}orique
\footnote{Laboratoire du CNRS, de l'Ecole Normale Sup\'{e}rieur
et de l'Universit\'e Paris-Sud.} , ENS 24 rue Lhomond}
\baselineskip=10pt
\centerline{\footnotesize\it F75231 Paris, France}
\vspace*{0.225truein}


\vspace*{0.21truein}
\abstracts{We study the decoherence process associated with 
the scattering of stochastic gravitational waves.
We discuss the case of macroscopic systems, such as the 
planetary motion of the Moon around the Earth, for which
gravitational scattering is found to dominate decoherence
though it has a negligible influence on damping. 
This contrast is due to the very high effective temperature 
of the  background of gravitational waves in our galactic environment.
}{}{}

\textlineskip			
\vspace*{12pt}			

\setcounter{footnote}0 
\renewcommand{\thefootnote}{\alph{footnote}}

\vspace*{1pt}\textlineskip	
\section{Decoherence and the micro/macro transition}	
\vspace*{-0.5pt}
\noindent
Decoherence is a general phenomenon which occurs in principle
for any physical system coupled to its environment. The fluctuations 
associated with this coupling tend to wash out quantum coherences -
{\it i.e.} superpositions of different quantum states - on a time 
scale which becomes extremely short for systems with a large degree 
of classicality - {\it i.e.} for superpositions of quantum states 
with sufficiently different classical properties 
\cite{Zeh70,Dekker77,Zurek81,Caldeira83,Joos85}. 

Decoherence thus plays an important role in the transition between 
microscopic and macroscopic physics.
For large macroscopic masses, say the Moon orbiting around the Earth,
decoherence is expected to be so efficient that the classical
description of the motion is sufficient. Precisely, the decoherence 
time scale is so short that the observation of any quantum coherence
effect is impossible. 
For microscopic masses in contrast, decoherence is expected to be so 
unefficient that we are left with the ordinary quantum description
of the system. In other words, if we consider for example an electron 
orbiting around a proton, the decoherence time scale is
so long that decoherence can be forgotten. 

A lot of theoretical papers have been devoted to decoherence
but only a few experiments have shown evidence
for the phenomenon and this can be understood from the 
simple arguments sketched in the previous paragraph.
In order to observe decoherence, one must deal with
`mesoscopic' systems for which the decoherence time is neither
too long nor too short. The transition from quantum to classical 
behaviour is then characterized through the variation of this
time with some parameter measuring the degree of classicality 
of the system. 

These experimental challenges have been met with a few specific
systems such as microwave photons stored in a high-Q 
cavity \cite{Brune96,Maitre97} or an ion in a trap \cite{Myatt00}.
In these model systems where the fluctuations are particularly
well mastered, the quantum/classical transition has been shown 
to fit the predictions of decoherence theory \cite{Raimond01}.

It must be emphasized that the details of the 
quantum/classical transition depend on the coupling mechanisms 
between the system under consideration and its environment
as well as on the spectral properties of the fluctuations.
Furthermore, a given system may be coupled to various 
environments which contribute differently to decoherence. 
It is only after having studied these points that one may obtain
a reliable estimation of the decoherence time scale and, hence, 
a precise description of the position of the frontier between
quantum and classical behaviours.

For motions in the solar system for instance, 
decoherence is often attributed to the scattering of the electromagnetic 
fluctuations associated with solar radiation or cosmic microwave background. 
In fact, as discussed here, the decoherence of planetary motions is not 
dominated by electromagnetic processes but rather by the scattering of 
stochastic gravitational waves present in our galactic environment.

\section{Decoherence and gravity}	
\vspace*{-0.5pt}
\noindent
The idea that the quantum/classical transition might be related to
fundamental fluctuations of space-time has often been proposed.
It can be presented in quite simple words relying essentially on
dimensional arguments. 

Fundamental fluctuations of space-time, associated for example with 
quantum gravity models, are expected to become important on length scales 
of the order of the Planck length, {\it i.e.} the length built on the 
constants $\hbar$, $c$ and $G$ 
\beq
&&\lP = \sqrt{\frac{\hbar G}{c^3}} \sim 10^{-35}\ \m
\eeq
This Planck length is extremely small when compared not only
to ordinary macroscopic scales but also to the smallest microscopic 
scales experimentally explored to date. 
The same argument holds for the Planck time 
$\tP = \frac \lP c$.

Meanwhile, the Planck scale for mass lies on the borderland 
between microscopic and macroscopic masses
\beq
&& \mP = \sqrt{\frac{\hbar c}{G}} \sim 22 \mu \g
\eeq
Hence, microscopic and macroscopic values of mass $m$ may be 
delineated by comparing the associated Compton length 
$\lC$ to the Planck length $\lP$ 
\beq
m < \mP &\Leftrightarrow & \lC = \frac{\hbar }{mc} > \lP \nonumber \\
m > \mP &\Leftrightarrow & \lC = \frac{\hbar }{mc} < \lP
\eeq
It is quite tempting to consider that this coincidence
is not just accidental but that it might be a consequence
of the existence of fundamental gravitational fluctuations.
The idea was already present in the Feynman lectures on
gravitation \cite{Feynman} and it was more thoroughly
developed and popularized by a number of authors, for
example \cite{Karolyhazy66,Diosi89,Penrose96}. 

However, the simple dimensional arguments given above are not by 
themselves sufficient to reach a precise conclusion.
As already stated, the description of the quantum/classical
transition for a given system must depend on the details of the 
coupling of this system to its environment as well as on the
noise spectrum characterizing the amplitude of fluctuations
at the frequencies of interest for the motion of the system.

The aim of this paper is to show that reliable conclusions
can be drawn at least for some well defined problems.
We will study the decoherence of planetary motions, such as the motion 
of the Moon around the Earth, associated with the scattering of the 
stochastic background of gravitational waves present in our celestial 
environment. We will estimate the decoherence rate and show that it is
dominated by this contribution of gravitational fluctuations 
\cite{Reynaud01}.

\section{Gravitational backgrounds}	
\vspace*{-0.5pt}
\noindent
We now explain how we describe the fundamental fluctuations 
of space-time and their effect on the motion of matter
although we don't have a complete theory of quantum gravity at our
disposal.

The basic idea is that the frequency range of interest,
which depends on the specific system of interest, lies 
in any case far below Planck frequency.
For the motion of the Moon for instance, this frequency
range lies from the orbital frequency, roughly speaking 1$\mu$Hz,
to frequencies larger by a few orders of magnitude.
At these frequencies, general relativity is known to be an
accurate description of gravitational phenomena, and this statement
is essentially independent of the modifications of the theory
which will have to take place when a complete theory of quantum 
gravity will be available \cite{Jaekel95}. 

Stated in different terms, general relativity is certainly not the
final word but it can be used quite safely as an effective theory
of gravitation at the frequencies involved in planetary motions.
We may then conclude that the fundamental space-time fluctuations
we have to consider are simply the gravitational waves which are
predicted by the linearized version of Einstein theory of gravity
\cite{Landau,Misner}. Precisely, the fundamental fluctuations in 
our gravitational environment are the stochastic backgrounds of 
gravitational waves which are thoroughly studied in relation with 
the ongoing experimental development of gravitational wave 
detectors \cite{Schutz99,Maggiore00}. 

The effect of gravitational perturbations may in principle be described
in a manifestly gauge-invariant manner. As soon as this is checked out,
a specific gauge can be chosen. Here the calculations are made simpler
by chosing the transverse traceless (TT) gauge with metric perturbations
differing from zero only for purely spatial components $h_{\rm ij}$
(i,j=1,2,3 stand for the spatial indices whereas 0 will represent the 
temporal index). 

Then the gravitational waves are conveniently described through a
mode decomposition 
\beq
&&h_{\rm ij}\left( x\right) = \int \frac{{\rm d} ^4 k}{\left( 2\pi \right) ^4}
\ h_{\rm ij}\left[ k\right] e ^{ -ik_\mu x^\mu }  \qquad
h_{\rm 00} = h_{\rm 0i} = 0
\eeq
Any Fourier component is a sum over the two circular polarizations $h^\pm$ 
\beq
&&h_{\rm ij} \left[ k\right] = 
\left( \frac {\varepsilon _{\rm i}^{+} \varepsilon _{\rm j}^{+}} {\sqrt{2}}
\right) ^{*} h^{+}\left[ k\right] +
\left( \frac {\varepsilon _{\rm i}^{-} \varepsilon _{\rm j}^{-}} {\sqrt{2}}
\right) ^{*} h^{-}\left[ k\right]  
\eeq
with the gravitational polarization tensors obtained as products of the
polarization vectors $\varepsilon _{\rm i}^{\pm}$ well-known from
electromagnetic theory.
The gravitational waves correspond to wavevectors $k$ lying on the light cone
($k^2=0$) and they are transverse with respect to this wavevector
($k_{\rm i} h_{\rm ij}=0$).

We will consider for simplicity the case of stationary, unpolarized and
isotropic backgounds. Such backgrounds correspond to correlation
functions which are completely characterized by the number $n _\gw $ 
of gravitons per mode 
\footnote{Taking into account quantum fluctuations of gravity, and not only
classical gravitational waves, one should replace $n _\gw $
by $\left( \frac 12 + n _\gw  \right)$ with the term $\frac 12$ 
representing vacuum fluctuations. These vacuum fluctuations 
have been shown to lead to ultimate fluctuations of geodesic 
distances of the order of Planck length \cite{Jaekel94}. 
Here we ignore this subtlety because the gravitational backgrounds
discussed below correspond to the classical limit where the number of 
gravitons per mode is extremely large $n _\gw  \gg 1$.}
\beq
&&\left\langle h^{+} \left[ k\right] h^{+} \left[ k'\right] \right\rangle 
= \left\langle h^{-} \left[ k\right] h^{-} \left[ k'\right] \right\rangle 
= \left( 2\pi \right) ^4 \delta ( k + k' ) \delta ( k^2 ) 
\ 32 \pi ^2 \frac{G}{c^3} \hbar n _\gw 
\eeq

We now rewrite the noise spectrum characterizing gravitational fluctuations
by performing a few formal transformations. First, backgrounds are usually
written in terms of one metric component, say $h \equiv h_{12}$, at a fixed
spatial point as a function of time.
The fluctuations of $h$ are described by a noise spectrum $C_{hh}$ obtained
by summing over the directions of momentum for a given value of frequency
\beq
&&\left\langle h \left( t \right) h \left( 0 \right) \right\rangle 
= \int \frac{{\rm d} \omega}{2\pi}
\ C_{hh}\left[ \omega \right] e ^{ -i\omega t }  
\eeq
$C_{hh}$ is the spectral density of strain fluctuations considered in most
papers on gravitational wave detectors \cite{Maggiore00}. 
We introduce an effective noise temperature $T _\gw $ for measuring the noise 
energy per mode and measure this temperature as a frequency $\Theta _\gw $
\beq
&& \kB T _\gw  \equiv  \hbar \omega n _\gw  
\equiv \frac{\hbar}{\pi} \ \Theta _\gw  
\eeq
with $\kB$ the Boltzmann constant. Then the spectral density $C_{hh}$ is 
simply the product of the frequency $\Theta _\gw $ by
the square of the Planck time $\tP $ 
\beq
&&C_{hh} \left[ \omega \right]  
= \frac{16 G}{5 c^5} \ \kB T _\gw  
= \frac{16}{5\pi} \ \Theta _\gw  \ \tP ^2
\eeq
$C_{hh}$ has the dimension of the inverse of a frequency.

It is worth stressing that $T _\gw $ is an effective noise 
temperature, that is an equivalent manner for representing 
the noise spectrum $C_{hh}$, but certainly not a real temperature. 
In particular, we have not supposed that $T _\gw $ is independent
of frequency. And, in any case, the value obtained below for this
temperature is much higher than the thermodynamical temperature 
associated with any known phenomenon.
This reveals that the motion of matter is so weakly
coupled to the gravitational fluctuations that it remains 
always far from the thermodynamical equilibrium. In other words,
the characteristic time which could be associated to the
potential thermalization would be extremely long.

We now use the information available from the 
studies devoted to the detectability of gravitational 
background by interferometers \cite{Schutz99,Maggiore00}. 
As already noticed, the orbital frequency of the Moon
is close to 1$\mu$Hz and the frequency range
of interest for our problem is roughly 1-100$\mu$Hz.
In this range, the gravitational background is dominated 
by the confusion noise due to gravitational waves emitted 
by unresolved binary systems in our galaxy and its
vicinity. Figure 1 of reference \cite{Schutz99} shows that this
`binary confusion background' corresponds to a nearly flat
function $C_{hh}$ in the frequency range of interest
\beq
10^{-6} \Hz < \frac \omega {2\pi} < 10^{-4}\Hz &\qquad&
C_{hh} \sim 10^{-34} \Hz^{-1}
\eeq
After the conversion already discussed, this corresponds 
to an equivalent noise temperature nearly constant on
the frequency range but with an extremely large 
\footnote{This temperature is even larger than Planck temperature
$\frac \hbar {\kB \tP} \sim 10^{32}\ \K$, which
emphasizes its unconventional character from the
point of view of thermodynamics.}
value $T _\gw  \sim 10^{41}\ \K$.
Equivalently, this corresponds to $\Theta _\gw  \sim 10^{52} \ \Hz$
or to a graviton number per mode 
$n_\gw \sim 10^{57}$ at $\omega \sim 1\mu$Hz. 
These numbers clearly correspond to the
high temperature limit $k_{\rm B}T_\gw \gg \hbar \omega $
where fluctuations may be described as classical.

It is worth noticing that these fluctuations are indeed
evaluated as the classical gravitational waves emitted by
binary systems in the galaxy or its vicinity.
They may be treated as stochastic variables because of
the large number of unresolved and independent sources.
As a consequence of the central limit theorem, they may
even be considered to obey a gaussian statistics, a 
property which will be used later on
\footnote{At this point, we may address an interesting objection
which states that, since these fluctuations are classical, they can 
be monitored and their effect corrected. 
This remark is valid as a matter of principle 
but irrelevant for the problem considered in this paper. 
Indeed, macroscopic bodies such as the Moon follow 
passively their geodesics and have therefore their motion affected 
by gravitational fluctuations. Only actively driven bodies,
such as dragfree satellites, have the ability to correct 
their trajectory from the effect of external forces \cite{Courty00}. 
And this ability may apply to the effect of gravitational 
waves only if the satellite can simultaneously monitor these waves~!} . 

The estimations discussed here correspond to the confusion background 
of gravitational waves emitted by binary systems in our galaxy or its
vicinity. They thus rely on the laws of physics and astrophysics 
as they are known in our local celestial environment. There also exist 
predictions for gravitational backgrounds associated with a variety of 
cosmic processes \cite{Maggiore00}. 
These predictions depend on the parameters used in the cosmic models
and they have a more speculative character than local astrophysical 
predictions. 
The associated temperatures vary quite rapidly with frequency and they 
are usually thought to be dominated by the confusion binary background 
in the $\mu$Hz frequency range. Should they surpass the binary confusion 
background, the latter would have to be considered as a minimum noise level 
in our gravitational environment and most conclusions to be drawn in the
following would be essentially preserved.

\section{Tidal perturbation of planetary motions}
\vspace*{-0.5pt}
\noindent
As the next step in our derivation, we now discuss the effect
of gravitational fluctuations on the motion of matter. As far as
the non relativistic limit of macroscopic motion with velocities
much smaller than the velocity of light is concerned, this effect
is essentially described by the standard theory of gravitational 
wave emission and gravitational wave detection by mechanical
detectors \cite{Landau,Misner}. 

In the non relativistic limit, the gravitational perturbation on a planetary
system may be represented as a tidal acceleration acting on each mass 
\beq
&&x_{\rm i}^{\prime \prime} \left( t\right)  = -R_{\rm i0j0} \left( t\right) 
x_{\rm j}\left( t\right) 
\eeq
The tidal tensor $R_{\rm i0j0}$ is built up from components of the Riemann 
curvature tensor and the prime denotes a time derivative. 
In the transverse traceless (TT) gauge, the tidal tensor is the second 
order derivative of the metric perturbation $h_{\rm ij}$ evaluated at 
the center of mass of the planetary system 
\beq
&&R_{\rm i0j0} \left( t\right) = -\frac{1}{2} h_{\rm ij}^{\prime \prime}
\left( t\right)
\eeq
The interaction is equivalently described by a perturbation
coupling the quadrupole momentum of the system to the tidal tensor.

We consider the simple case of a circular planetary orbit
in the plane $x_{\rm 1}x_{\rm 2}$.  
The two masses $m_{a}$ and $m_{b}$ are also described by
the reduced mass $m=\frac{m_{a}m_{b}}{m_{a}+m_{b}}$ and
the total mass $M=m_{a}+m_{b}$. The radius $\rho$, that is the constant
distance between the two masses, and the orbital frequency $\Omega$
are related by the third Kepler law 
\beq
&&\rho ^{3}\Omega ^{2} = GM
\eeq
In the following we will also use as characteristic parameters the 
tangential velocity $v=\rho \Omega$ and the normal acceleration 
$a = \rho \Omega ^2 = \frac {v ^2} \rho$.

In this simple configuration, we characterize the gravitational 
perturbation through a relative force $F$ projected along the mean motion 
\beq
&&F\left( t\right)  = \frac{m\rho } {2\sqrt{2}} 
\left( h^{\prime \prime} \left( t\right) e^{ 2i\Omega t} +
{\rm c.c.} \right) \qquad
h =\frac {h_{\rm 12}} {\sqrt{2}} + \frac{h_{\rm 22}-h_{\rm 11}}{2i\sqrt{2}}
\eeq
$F$ is written in terms of the circular polarization $h$ which fits the 
circular motion of the planetary system in the plane $x_{\rm 1}x_{\rm 2}$
of the orbit. 
Precisely, the force $F$ is driven by gravitational waves through a 
frequency transposition due to the evolution of the quadrupole momentum 
at the frequency $2\Omega $. 

The effect of this force on motion is essentially a momentum diffusion.
As a matter of fact, if we define the momentum perturbation $p_t $ 
integrated over an interaction time $t$, we obtain a result typical
of Brownian motion with the variance of $p_t$ proportional to the
elapsed time $t$  
\beq
&&p_t  = \int_0 ^t {\rm d}s\ F\left( s\right) \qquad
\Delta p_t^2 \equiv \left\langle p_t ^{2}\right\rangle = 2 D_\gw t
\eeq
The momentum diffusion coefficient $D_\gw$ is obtained as 
\cite{Reynaud01}
\beq
&&2D_\gw = 4m^{2}a^{2}C_{hh}\left[ 2\Omega \right]   
\eeq
with $a$ the acceleration and $C_{hh}$ the gravitational noise at the 
frequency $2\Omega $ of evolution of the quadrupole 
\footnote{We have used the assumption of an unpolarized and isotropic background,
which seems reasonnable for a background of extragalactic origin
but not necessarily for a background of galactic origin.
The generalization of this result to a non isotropic
background would however not modify strongly the orders of magnitude
discussed in the following} .

The diffusion coefficient may be written under the form of an Einstein 
fluctuation-dissipation relation, i.e. as the product of the effective 
temperature $T_\gw$ by a damping rate associated with the emission of 
gravitational waves by the planetary system 
\beq
&&D_\gw = m \Gamma _\gw \kB T_\gw \qquad \Gamma _\gw 
=\frac{32Gma^{2}}{5c^{5}}  
\eeq
This formula connects the Einstein fluctuation-dissipation 
relation on Brownian motion \cite{Einstein05} and the Einstein quadrupole
formula for gravitational wave emission \cite{Einstein18}.
The latter does not depend on the effective temperature. For the Moon, 
the associated damping is so small, $\sim 10^{-34}{\rm s}^{-1}$, 
that it does not affect the classical motion. 
It is only for strongly bound binary systems that gravitational damping 
has a noticeable effect \cite{Taylor92}. 
In contrast, the damping rate $\Gamma _\gw$ vanishes at the limit of a 
null acceleration, that is also at the limit of an inertial motion. 

\section{Decoherence of planetary motions}
\vspace*{-0.5pt}
\noindent
We come now to the final step of our evaluation of decoherence
for planetary motions. To this aim, we consider two neighbouring 
motions on the circular orbit of the Moon around the Earth.
More precisely, we consider two motions characterized by the same 
spatial geometry but slightly different values of the epoch
- {\it i.e.} the time of passage at a given space point.
For simplicity, the difference is measured by the spatial distance $\Delta x$ 
between the two motions which is constant on a circular orbit.

As the gravitational wave fluctuations depend on time,
these two motions undergo different perturbations. 
This effect is conveniently described by the differential 
perturbation $\delta S_t $ on the action integral 
after an interaction time $t$. In a first order
perturbation theory, $\delta S_t $ is simply evaluated as 
\beq
&&\delta S_t = \int_0 ^t {\rm d}s F\left( s\right) \Delta x 
= p_t  \Delta x
\eeq
where $p_t$ is the momentum perturbation integrated over $t$ for
a single motion.

Should we associate a quantum phase $\Phi$ to a motion of the Moon, 
the two neighbouring motions would suffer a differential dephasing
characterized by the exponential 
\beq
&& e ^{i \delta \Phi_t} = \exp \frac{i \delta S_t }{\hbar } 
\eeq
Since $\delta \Phi_t $ is linearly driven by gravitational waves
behaving as classical fluctuations, we can treat it as a gaussian 
classical stochastic variable and this allows us to get the mean value 
of the exponentiated dephasing as 
\beq
&&\left\langle e ^{i \delta \Phi_t} \right\rangle 
= \exp \left( -\frac {\Delta \Phi_t^2} 2 \right) 
\qquad \Delta \Phi_t^2 \equiv \left\langle \delta \Phi_t^2 \right\rangle 
= \frac {\Delta S_t^2}{\hbar^2} 
\eeq

Using the Einstein relation for momentum diffusion, we obtain 
the final characterization 
of decoherence between two neighbouring trajectories 
\beq
&&\left\langle e ^{i \delta \Phi_t} \right\rangle =
\exp \left( -\Lambda _\gw  \Delta x^2 t\right)   \nonumber \\
&&\Lambda _\gw = \frac{D_\gw }{\hbar^2}
=\frac{32Gm^2 a^2}{5c^5 \hbar^2} \kB T_\gw   
\eeq
As expected from general discussions \cite{Zurek81}, decoherence is 
described by a factor decreasing exponentially with time $t$, the 
inverse $\Lambda _\gw \Delta x^2 $ of the decoherence time becoming 
larger when the distance $\Delta x$ increases.
For the motion of the Moon, the coefficient $\Lambda _\gw $ 
is so large $\sim 10^{75}{\rm s}^{-1}{\rm m}^{-2}$ 
that an extremely short decoherence time is obtained even for
ultrasmall distances $\Delta x$. To fix ideas, this time
lies in the $10\mu$s range for $\Delta x$ 
of the order of the Planck length.

\section{Discussion}	
\vspace*{-0.5pt}
\noindent
This result confirms the idea that decoherence is so efficient for large 
macroscopic masses that their motion can be treated as classical.
But it also leads to more specific conclusions.

First, we want to come back to the simple scaling arguments presented
in the introduction. To this aim, we rewrite the decoherence factor 
\beq
&&\left\langle e ^{i \delta \Phi_t} \right\rangle =
\exp \left( - \frac{32}5 \left( \frac{m v^2}{\mP c^2} \right) ^2 
\left( \frac {\Delta x} \rho \right) ^2 \Theta_\gw t \right) 
\eeq
The ratio $\frac{m^2}{\mP^2}$ which appears in this expression is clearly 
reminiscent of the scaling arguments showing the role of the Planck mass 
$\mP$ as a natural reference on the borderland between microscopic and 
macroscopic masses. However the presence of the other terms in the formula 
implies that these arguments are not sufficient for obtaining
quantitative estimations. The correct result depends on the velocity 
\footnote{In fact, the decoherence factor scales as the square of the
kinetic energy on one hand and on the elapsed time on another hand.} ,
on the geometrical factor $\frac {\Delta x} \rho$
and on the frequency $\Theta _\gw$ which measures the amplitude of 
gravitational fluctuations at the frequency of interest for the
motion under study. 

Then, the gravitational contribution to decoherence studied in this paper 
is found to overshadow the other contributions which dominate
the damping and are for this reason usually studied.
The damping of the main motion of Moon, revealed by the secular variation 
of lunar rotation \cite{Bois96}, is due to the interaction of Earth and 
Moon tides and the corresponding damping rate is $\sim 10^{16}$ larger 
than the gravitation contribution $\Gamma_\gw$. The effect of radiation pressure
of solar photons is $\sim 10^{10}$ larger than $\Gamma_\gw$ and even the scattering
of the cosmic microwave background would dominate $\Gamma_\gw$ by a factor larger
than 100 for the case of Moon. 
At the same time however, the effective gravitational temperature has such
a large value that the gravitational decoherence coefficient $\Lambda_\gw$ 
is much larger than the coefficients associated with tide interactions and
electromagnetic scattering processes. 

This entails that the ultimate fluctuations of the motion of Moon, and the
associated decoherence mechanisms, are determined by the classical gravitation 
theory which also explains its mean motion. 
In other words, the environment to be considered when dealing with
macroscopic motion consists in the gravitational waves of the confusion
binary background.
This background is naturally defined in the reference frame of the galaxy
if it is dominated by galactic contributions or in a reference frame
built on a larger region of the universe if extragalactic contributions
have to be taken into account.
In any case, the gravitational background allows to define a reference
frame built on our celestial environment which has an effect on local 
phenomena. 

\nonumsection{Acknowledgements}
\noindent
P.A.M.N. wishes to thank CAPES, CNPq,
FAPERJ, PRONEX, COFECUB, ENS and MENRT for their financial support which
made possible his stays in Paris during which this work was performed.

\eject

\end{document}